\documentclass[
showpacs,
floatfix,
aps,
prl,
twocolumn,
superscriptaddress,
amssymb,
]{revtex4-1}

\usepackage{amssymb}
\usepackage{latexsym}
\usepackage[dvips]{graphicx}
\usepackage{amsmath}

\usepackage{graphicx}
\usepackage{dcolumn}
\usepackage{amsfonts}
\usepackage{bm}
\usepackage{epsfig}

\newcommand{\be}{\begin{equation}}
\newcommand{\ee}{\end{equation}}
\newcommand{\bea}{\begin{eqnarray}}
\newcommand{\eea}{\end{eqnarray}}

\def\bip{B_\parallel}

\def\A{A_{R}}

\def\rxx{R_{xx}}
\def\ryy{R_{yy}}

\def\easy{\left < 110 \right >}

\def\x{\hat{x}}
\def\y{\hat{y}}

\def\a{\theta = 0^\circ}

\newcommand{\rfig}[1]{Fig.\,\ref{#1}}
\newcommand{\rFig}[1]{Figure \,\ref{#1}}

\begin{document}
\title{Evidence for a new symmetry breaking mechanism reorienting quantum Hall nematics}
\author{Q. Shi}
\affiliation{School of Physics and Astronomy, University of Minnesota, Minneapolis, Minnesota 55455, USA}
\author{M. A. Zudov}
\email[Corresponding author: ]{zudov@physics.umn.edu}
\affiliation{School of Physics and Astronomy, University of Minnesota, Minneapolis, Minnesota 55455, USA}
\author{J. D. Watson}
\affiliation{Department of Physics and Astronomy, Purdue University, West Lafayette, Indiana 47907, USA}
\affiliation{Birck Nanotechnology Center, Purdue University, West Lafayette, Indiana 47907, USA}
\author{G. C. Gardner}
\affiliation{Birck Nanotechnology Center, Purdue University, West Lafayette, Indiana 47907, USA}
\affiliation{School of Materials Engineering, Purdue University, West Lafayette, Indiana 47907, USA}
\author{M. J. Manfra}
\affiliation{Department of Physics and Astronomy, Purdue University, West Lafayette, Indiana 47907, USA}
\affiliation{Birck Nanotechnology Center, Purdue University, West Lafayette, Indiana 47907, USA}
\affiliation{School of Materials Engineering, Purdue University, West Lafayette, Indiana 47907, USA}
\affiliation{School of Electrical and Computer Engineering, Purdue University, West Lafayette, Indiana 47907, USA}
\begin{abstract}

We report on the effect of in-plane magnetic field $\bip$ on stripe phases in higher ($N=2,3$) Landau levels of a high-mobility 2D electron gas.
In accord with previous studies, we find that a modest $\bip$ applied parallel to the native stripes aligns them perpendicular to it. 
However, upon further increase of $\bip$, stripes are reoriented back to their native direction.
Remarkably, applying $\bip$ perpendicular to the native stripes also aligns stripes parallel to it.
Thus, regardless of the initial orientation of stripes with respect to $\bip$, stripes are ultimately aligned \emph{parallel} to $\bip$. 
These findings provide evidence for a $\bip$-induced symmetry breaking mechanism which challenge current understanding of the role of $\bip$ and should be taken into account when determining the strength of the native symmetry breaking potential.
Finally, our results might indicate nontrivial coupling between the native and external symmetry breaking fields, which has not yet been theoretically considered.

\end{abstract}
\maketitle

Electronic liquid crystal-like phases, termed electron nematics or stripes, are expected to form in a wide variety of condensed matter systems \cite{fradkin:2010,borzi:2007,daou:2010,chu:2010,okazaki:2011}. 
A two-dimensional electron gas in GaAs/AlGaAs hosts the first, and perhaps the most striking, realization of such phases \cite{koulakov:1996,fogler:1996,lilly:1999a,du:1999,fradkin:1999,fradkin:2000}.
Stripes in a two-dimensional electron gas form due to interplay between exchange and direct Coulomb interactions \cite{koulakov:1996,fogler:1996,fradkin:1999,fradkin:2000} and are manifested by the resistivity minima (maxima) in the easy (hard) transport direction near half-integer filling factors, $\nu=9/2,11/2,13/2,...$ when the system is cooled below $T \approx 0.1$ K. 
With very few exceptions \cite{zhu:2002,pollanen:2015,liu:2013}, stripes in GaAs are aligned along $\easy$ direction, but what exactly causes such orientation remains unknown \cite{sodemann:2013,kovudayur:2011,pollanen:2015}. 

While the origin of the native symmetry-breaking potential responsible for preferred stripes orientation remains elusive, its magnitude was routinely obtained from experiments employing in-plane magnetic field $\bip$ which provides an external symmetry-breaking field competing with and overcoming the native one.
Our current understanding of $\bip$-induced symmetry-breaking potential is based on finite thickness effects \cite{jungwirth:1999,stanescu:2000}, which favor stripes perpendicular to $\bip$, consistent with previous experiments \cite{lilly:1999b,pan:1999,cooper:2001,zhu:2009,pollanen:2015,shi:2016b}.
The same approach successfully explains $\bip$-induced stripes in both single-subband \cite{lilly:1999b,pan:1999,xia:2010,xia:2011,friess:2014} and double-subband \cite{pan:2000} systems.

In this Rapid Communication we re-examine the effect of in-plane magnetic field on quantum Hall stripes in ultrahigh quality GaAs quantum wells.
In agreement with early experiments \cite{lilly:1999b,pan:1999,cooper:2001}, we find that a $\bip \lesssim 0.4$ applied parallel to the native stripes aligns stripes perpendicular to it.
Remarkably, upon further increase of $\bip$, stripes are reoriented back to their native direction, i.e. \emph{parallel} to $\bip$.
When $\bip$ is applied perpendicular to the native stripes, we also find that stripes are reoriented \emph{parallel} to $\bip$.
We thus conclude that there exist a new $\bip$-induced symmetry-breaking potential which challenge our understanding of the role of $\bip$ and must be taken into account when determining the strength of the native symmetry-breaking potential.
In contrast to the well-established $\bip$-induced symmetry-breaking potential originating from finite thickness effects, new symmetry-breaking potential exhibits high sensitivity to spin and Landau level indices.
Finally, our results suggest a nontrivial coupling between the native and external symmetry-breaking fields, which has not yet been theoretically considered and might provide an important clue to unveiling the origin of the native symmetry-breaking potential.

The sample used in our study is a $4\times4$ mm square cleaved from a symmetrically doped, 30 nm-wide GaAs/AlGaAs quantum well. 
Electron density and mobility were $n_e \approx 2.9 \times 10^{11}$ cm$^{-2}$ and $\mu \approx 1.6 \times 10^7$ cm$^2$/Vs, respectively.
Eight indium contacts were fabricated at the corners and midsides of the sample. 
The longitudinal resistances, $\rxx$ and $\ryy$, were measured using four-terminal, low-frequency lock-in technique; the current (typically 10 nA) was sent through the midside contacts and the voltage drop was recorded between the corner contacts.
An in-plane magnetic field was introduced by tilting the sample about $\x$ or $\y$ axis, in two separate cooldowns.
The data were taken at $T \approx 20$ mK.

%%%%%%%%%%%%%%%%%%%%%%%%%%%%%%%%%%
\begin{figure}[t]
\includegraphics{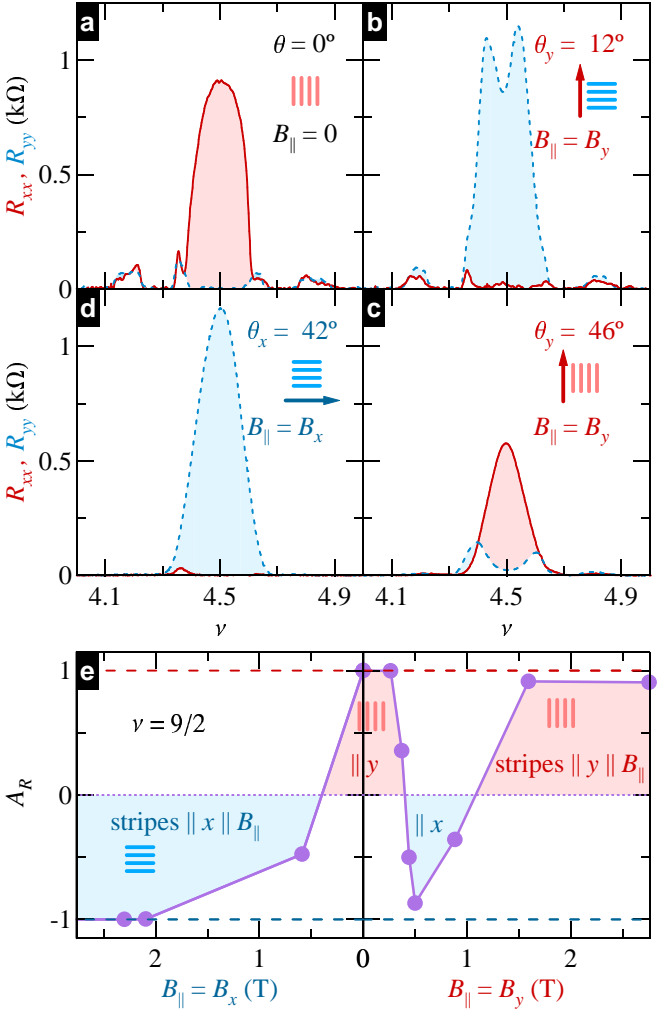}
\caption{
$\rxx$ (solid line) and $\ryy$ (dotted line) versus $\nu$ at (a) $\a$, (b) $\bip = B_y$ and $\theta_y = 12^{\circ}$, (c) $\bip = B_y$ and $\theta_y = 46^{\circ}$, and (d) $\bip = B_x$ and $\theta_x = 42^{\circ}$.
(e) Resistance anisotropy $\A \equiv (\rxx-\ryy)/(\rxx+\ryy)$ as a function of $\bip = B_x$ (left) and $\bip = B_y$ (right) at $\nu$ = 9/2.
The inset shows the stripes orientation and the direction of $\bip$.
}
\label{fig1}
\end{figure}
%%%%%%%%%%%%%%%%%%%%%%%%%%%%%%%%%%

In \rfig{fig1}(a) we present an example of stripes in perpendicular magnetic field near $\nu = 9/2$, characterized by $\rxx \gg \ryy$, indicating stripes oriented along $\y \equiv \easy$ direction.
When $\bip$ is applied parallel to the native stripes ($\bip = B_y$), at $\theta_y = 12^{\circ}$ stripes reorient along $\x$-direction (perpendicular to $\bip$), as anticipated [see \rfig{fig1}(b)].
Surprisingly, upon further increase of $B_y$, at $\theta_y = 46^{\circ}$ stripes are reoriented again, back to their native direction and are now aligned \emph{parallel} to $\bip$ [see \rfig{fig1}(c)].
When $\bip$ is applied perpendicular to the native stripes ($\bip = B_x$), at $\theta_x = 42^{\circ}$ stripes are reoriented along $\x$-direction, again aligning \emph{parallel} to $\bip$ [see \rfig{fig1}(d)].
We thus conclude that, regardless of the orientation of $\bip$, ultimately, stripes align \emph{parallel} to $\bip$.

\rFig{fig1}(e) shows the resistance anisotropy $\A \equiv (\rxx-\ryy)/(\rxx+\ryy)$ vs $B_x$ (left panel) and $B_y$ (right panel).
Starting from $\A \approx$ 1, with increasing $B_y$, $\A$ vanishes at $B_y \approx 0.4$ T, reaches $\A \approx -1$, turns around, disappears again at $B_y \approx 1.1$ T, and finally returns to $\A \approx$ 1.
With increasing $B_x$, $\A$ vanishes at $B_x \approx 0.4$ T and then approaches $\A \approx -1$.
Taken together, the data in \rfig{fig1} clearly demonstrate the existence of a mechanism which favors stripes along $\bip$.
As we show next, this mechanism is relevant at other filling factors although it shows sensitivity to both the spin index $\sigma$ and the Landau level index $N$.

%%%%%%%%%%%%%%%%%%%%%%%%%%%%%%%%%%
\begin{figure}[t]
\includegraphics{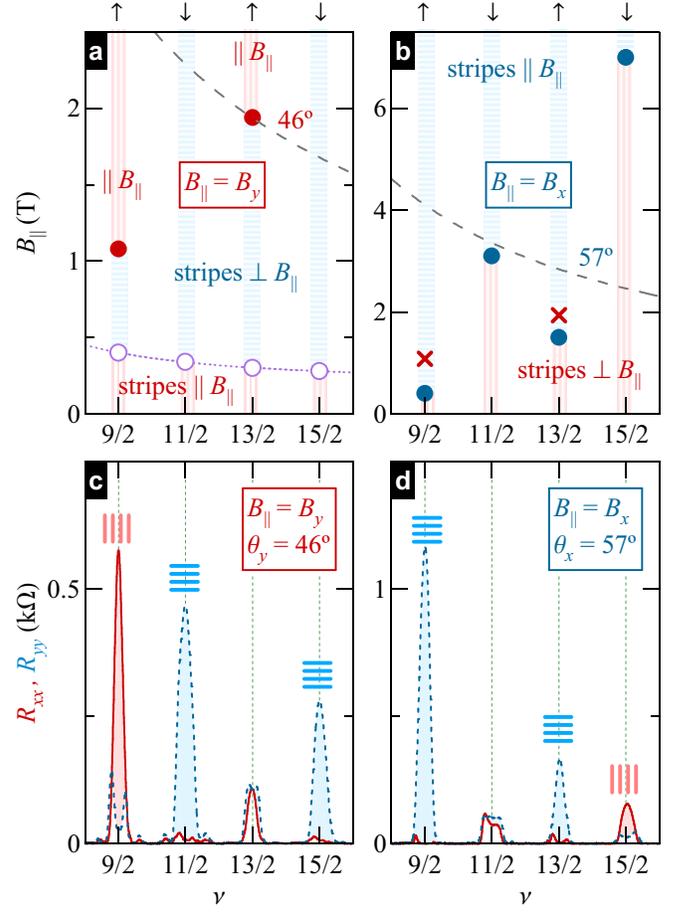}
\caption{
Evolution of stripe orientation at $\nu$ = 9/2, 11/2, 13/2 and 15/2 for (a) $\bip = B_y$ and (b) $\bip = B_x$. 
The regions with vertical (horizontal) lines represent stripes along $\y$ ($\x$), demarcated by the characteristic fields $B_{1y}$ (open dots), $B_{2x}$ [panel (b)] and $B_{2y}$ [panel (a)] (solid dots).
For comparison, $B_{2y}$ (crosses) are added to panel (b) at $\nu = 9/2$ and $13/2$.
Dashed lines represent $\bip$ at marked tilt angles.
$\rxx$ (solid line) and $\ryy$ (dotted line) vs $\nu$ at (c) $\theta_y = 46^\circ$ and (d) $\theta_x = 57^\circ$.
}
\label{fig2}
\end{figure}
%%%%%%%%%%%%%%%%%%%%%%%%%%%%%%%%%%

\rFig{fig2} shows the phase diagram of stripe orientations at $\nu$ = 9/2, 11/2, 13/2 and 15/2 for (a) $\bip = B_y$ and (b) $\bip = B_x$. 
The regions representing stripes along $\y$ ($\x$), marked by vertical (horizontal) lines, are demarcated by the characteristic fields (circles) where $\A \approx 0$.
We define $B_{1y}$ as the field at which stripes reorient perpendicular to $B_y$, and $B_{2x}$ ($B_{2y}$) as the field at which stripes reorient parallel to $B_x$ ($B_y$). 
The dashed lines represent $\bip$ at select tilt angles and the resistances at these angles are shown in \rfig{fig2}(c) and (d).
First, we notice that both $B_{2x}$ and $B_{2y}$ increase with $N$ for a given $\sigma$ while $B_{1y}$ decreases with $N$.
Second, strong sensitivity to $\sigma$ renders the overall dependencies of $B_{2x}$ and $B_{2y}$ on $\nu$ nonmonotonic;
indeed, both $B_{2x}$ and $B_{2y}$ are always considerably smaller for $\sigma = +1/2$ than $\sigma = -1/2$.

The sensitivity of $B_{2x}$ and $B_{2y}$ to $\sigma$ and $N$ is also evident in the raw data presented in \rfig{fig2}(c) and (d).
For $\bip = B_y$ and $\theta_y = 46^\circ$ [see \rfig{fig2}(c)], the stripes at $\nu = 9/2$ have reoriented back to their native direction (along $\y$ axis) as manifested by $\rxx \gg \ryy$.
The data at $\nu =13/2$ suggest that the (second) reorientation is about to happen at this filling factor as well. 
However, stripes at $\nu = 11/2$ and 15/2 are still oriented along $\x$ direction, as $\rxx \ll \ryy$.
For $\bip = B_x$ and $\theta_x = 57^\circ$ [see \rfig{fig2}(d)], stripes both at $\nu = 9/2$ and 13/2 have reoriented along $\x$ axis, stripes at $\nu = 11/2$ are undergoing the reorientation, while stripes at $\nu = 15/2$ are still oriented along $\y$ direction.

To compare the magnitudes of $B_{2x}$ and $B_{2y}$ we add $B_{2y}$ (crosses) to \rfig{fig2}(b) and observe that $B_{2y}$ is close to $B_{2x}$ at both $\nu$ = 9/2 and 13/2.
Combined with qualitatively identical dependence on $\sigma$ and $N$, this observation suggests that the reorientations characterized by $B_{2x}$ and $B_{2y}$ are of similar origin.
We can now classify the reorientations into two types.
The first type, which aligns stripes perpendicular to $\bip$, is characterized by $B_{1y}$ that is not sensitive to $\sigma$ and decreases with $N$.
The second type aligns stripes parallel to $\bip$ and is characterized by $B_{2x}$ and $B_{2y}$ that depend on $\sigma$ \cite{note:90} and increase with $N$.
The data at $\nu = 9/2$ also suggest that the mechanism responsible for reorientation of the first (second) type dominates at lower (higher) $\bip$.

The reorientation of the first type is well understood in terms of finite thickness effects \cite{jungwirth:1999,stanescu:2000}.
The $\bip$-induced anisotropy energy can be defined as $E_{1A} = E_{1\parallel} - E_{1\perp}$ , where $E_{1\parallel}$ and $E_{1\perp}$ are optimized energies per electron of a stripe state parallel and perpendicular to $\bip$, respectively.
For a single-subband system, $E_{1A}>0$ and increases monotonically with $\bip$ \cite{jungwirth:1999,cooper:2001}. 
While $E_{1A}$ could change sign in systems with two occupied subbands, such systems do not exhibit native stripes at $\bip = 0$ and $E_{1A}$ is insensitive to $\sigma$ \cite{jungwirth:1999,pan:2000}.
We thus conclude that reorientations of the second type, favoring stripes parallel to $\bip$, have a distinct physical origin.

Furthermore, since $E_{1A}(B_x) = E_{1A}(B_y)$ for $B_x = B_y$ \cite{jungwirth:1999,stanescu:2000}, we can conclude that the mechanism responsible for reorientation of the second type lacks such symmetry.
Indeed, if we assume that the anisotropy energy due to the second mechanism, $E_{2A}$, is the same for $\bip$ applied along $\x$ or $\y$ directions, one would expect, in the absence of native symmetry-breaking potential, $B_{2x} = B_{2y}$ at a given $\nu$; a native symmetry-breaking potential favoring stripes along $\y$ direction would then lead to $B_{2x} > B_{2y}$.
In contrast, our data show exactly the opposite, and, e.g., at $\nu$ = 9/2, $B_{2y} \approx$ 1.0 T is considerably larger than $B_{2x} \approx$ 0.4 T.
We thus conclude that $E_{2A}$ must depend on the direction of $\bip$, suggesting possible coupling of $\bip$ to native symmetry-breaking potential \cite{note:8}.
A proposal considering a combination of Rashba and Dresselhaus spin-orbital interactions as the origin of the native symmetry-breaking potential \cite{sodemann:2013,note:9,note:15} seems to indicate that the effects of $\bip$ on such native stripe states should be sensitive to its orientation with respect to the crystal axes.
However, a study of the interplay between $\bip$ and spin-orbital interactions was left for future work.

Despite our lack of understanding of the mechanism responsible for the second type of reorientation, our experimental results unambiguously demonstrate that two competing mechanisms must be incorporated in any complete theory of reorientation of quantum Hall nematics. 
Another important implication of our findings is related to the identification of the native symmetry-breaking potential, whose strength was traditionally obtained by calculating $E_{1A}$ at $\bip=B_{1c}$.
In light of clear evidence for the second mechanism and its possible coupling to the native symmetry-breaking potential and/or $E_{1A}$, this approach must be reexamined.

We thank G. Csathy, L. Engel, Y. Liu, B. Skinner, M. Shayegan, B. Shklovskii, and I. Sodemann for discussions.
We thank G. Jones, S. Hannas, T. Murphy, J. Park, and A. Suslov for technical support.
The work at Minnesota (Purdue) was supported by the U.S. Department of Energy, Office of Science, Basic Energy Sciences, under Award \# ER 46640-SC0002567 (DE-SC0006671).
Q.S. acknowledges The University of Minnesota Doctoral Dissertation Fellowship.
Experiments were performed at the National High Magnetic Field Laboratory, which is supported by NSF Cooperative Agreement No. DMR-0654118, by the State of Florida, and by the DOE.

%\bibliographystyle{apsrev-titles}
%\bibliography{../../bibRMP1qs.1_2,footnotes2}

\begin{thebibliography}{31}
\expandafter\ifx\csname natexlab\endcsname\relax\def\natexlab#1{#1}\fi
\expandafter\ifx\csname bibnamefont\endcsname\relax
 \def\bibnamefont#1{#1}\fi
\expandafter\ifx\csname bibfnamefont\endcsname\relax
 \def\bibfnamefont#1{#1}\fi
\expandafter\ifx\csname citenamefont\endcsname\relax
 \def \citenamefont#1{#1}\fi
\expandafter\ifx\csname url\endcsname\relax
 \def\url#1{\texttt{#1}}\fi
\expandafter\ifx\csname urlprefix\endcsname\relax\def\urlprefix{URL }\fi
\providecommand{\bibinfo}[2]{#2}
\providecommand{\eprint}[2][]{\url{#2}}

\bibitem[{ \citenamefont{Fradkin et~al.}(2010) \citenamefont{Fradkin, Kivelson,
 Lawler, Eisenstein, and Mackenzie}}]{fradkin:2010}
\bibinfo{author}{\bibfnamefont{E.}~\bibnamefont{Fradkin}},
 \bibinfo{author}{\bibfnamefont{S.~A.} \bibnamefont{Kivelson}},
 \bibinfo{author}{\bibfnamefont{M.~J.} \bibnamefont{Lawler}},
 \bibinfo{author}{\bibfnamefont{J.~P.} \bibnamefont{Eisenstein}},
 \bibnamefont{and} \bibinfo{author}{\bibfnamefont{A.~P.}
 \bibnamefont{Mackenzie}}, \emph{\bibinfo{title}{Nematic fermi fluids in
 condensed matter physics}}, \bibinfo{journal}{Annu. Rev. Condens. Matter
 Phys.} \textbf{\bibinfo{volume}{1}}, \bibinfo{pages}{153}
 (\bibinfo{year}{2010}).

\bibitem[{ \citenamefont{Borzi et~al.}(2007) \citenamefont{Borzi, Grigera,
 Farrell, Perry, Lister, Lee, Tennant, Maeno, and Mackenzie}}]{borzi:2007}
\bibinfo{author}{\bibfnamefont{R.~A.} \bibnamefont{Borzi}},
 \bibinfo{author}{\bibfnamefont{S.~A.} \bibnamefont{Grigera}},
 \bibinfo{author}{\bibfnamefont{J.}~\bibnamefont{Farrell}},
 \bibinfo{author}{\bibfnamefont{R.~S.} \bibnamefont{Perry}},
 \bibinfo{author}{\bibfnamefont{S.~J.~S.} \bibnamefont{Lister}},
 \bibinfo{author}{\bibfnamefont{S.~L.} \bibnamefont{Lee}},
 \bibinfo{author}{\bibfnamefont{D.~A.} \bibnamefont{Tennant}},
 \bibinfo{author}{\bibfnamefont{Y.}~\bibnamefont{Maeno}}, \bibnamefont{and}
 \bibinfo{author}{\bibfnamefont{A.~P.} \bibnamefont{Mackenzie}},
 \emph{\bibinfo{title}{Formation of a nematic fluid at high fields in
 Sr$_3$Ru$_2$O$_7$}}, \bibinfo{journal}{Science}
 \textbf{\bibinfo{volume}{315}}, \bibinfo{pages}{214} (\bibinfo{year}{2007}).

\bibitem[{ \citenamefont{Daou et~al.}(2010) \citenamefont{Daou, Chang, LeBoeuf,
 Cyr-Choiniere, Laliberte, Doiron-Leyraud, Ramshaw, Liang, Bonn, Hardy
 et~al.}}]{daou:2010}
\bibinfo{author}{\bibfnamefont{R.}~\bibnamefont{Daou}},
 \bibinfo{author}{\bibfnamefont{J.}~\bibnamefont{Chang}},
 \bibinfo{author}{\bibfnamefont{D.}~\bibnamefont{LeBoeuf}},
 \bibinfo{author}{\bibfnamefont{O.}~\bibnamefont{Cyr-Choiniere}},
 \bibinfo{author}{\bibfnamefont{F.}~\bibnamefont{Laliberte}},
 \bibinfo{author}{\bibfnamefont{N.}~\bibnamefont{Doiron-Leyraud}},
 \bibinfo{author}{\bibfnamefont{B.~J.} \bibnamefont{Ramshaw}},
 \bibinfo{author}{\bibfnamefont{R.}~\bibnamefont{Liang}},
 \bibinfo{author}{\bibfnamefont{D.~A.} \bibnamefont{Bonn}},
 \bibinfo{author}{\bibfnamefont{W.~N.} \bibnamefont{Hardy}},
 \bibnamefont{et~al.}, \emph{\bibinfo{title}{Broken rotational symmetry in the
 pseudogap phase of a high-$T_c$ superconductor}}, \bibinfo{journal}{Nature}
 \textbf{\bibinfo{volume}{463}}, \bibinfo{pages}{519} (\bibinfo{year}{2010}).

\bibitem[{ \citenamefont{Chu et~al.}(2010) \citenamefont{Chu, Analytis, De~Greve,
 McMahon, Islam, Yamamoto, and Fisher}}]{chu:2010}
\bibinfo{author}{\bibfnamefont{J.-H.} \bibnamefont{Chu}},
 \bibinfo{author}{\bibfnamefont{J.~G.} \bibnamefont{Analytis}},
 \bibinfo{author}{\bibfnamefont{K.}~\bibnamefont{De~Greve}},
 \bibinfo{author}{\bibfnamefont{P.~L.} \bibnamefont{McMahon}},
 \bibinfo{author}{\bibfnamefont{Z.}~\bibnamefont{Islam}},
 \bibinfo{author}{\bibfnamefont{Y.}~\bibnamefont{Yamamoto}}, \bibnamefont{and}
 \bibinfo{author}{\bibfnamefont{I.~R.} \bibnamefont{Fisher}},
 \emph{\bibinfo{title}{In-plane resistivity anisotropy in an underdoped iron
 arsenide superconductor}}, \bibinfo{journal}{Science}
 \textbf{\bibinfo{volume}{329}}, \bibinfo{pages}{824} (\bibinfo{year}{2010}).

\bibitem[{ \citenamefont{Okazaki et~al.}(2011) \citenamefont{Okazaki, Shibauchi,
 Shi, Haga, Matsuda, Yamamoto, Onuki, Ikeda, and Matsuda}}]{okazaki:2011}
\bibinfo{author}{\bibfnamefont{R.}~\bibnamefont{Okazaki}},
 \bibinfo{author}{\bibfnamefont{T.}~\bibnamefont{Shibauchi}},
 \bibinfo{author}{\bibfnamefont{H.~J.} \bibnamefont{Shi}},
 \bibinfo{author}{\bibfnamefont{Y.}~\bibnamefont{Haga}},
 \bibinfo{author}{\bibfnamefont{T.~D.} \bibnamefont{Matsuda}},
 \bibinfo{author}{\bibfnamefont{E.}~\bibnamefont{Yamamoto}},
 \bibinfo{author}{\bibfnamefont{Y.}~\bibnamefont{Onuki}},
 \bibinfo{author}{\bibfnamefont{H.}~\bibnamefont{Ikeda}}, \bibnamefont{and}
 \bibinfo{author}{\bibfnamefont{Y.}~\bibnamefont{Matsuda}},
 \emph{\bibinfo{title}{Rotational symmetry breaking in the hidden-order phase
 of URu$_2$Si$_2$}}, \bibinfo{journal}{Science}
 \textbf{\bibinfo{volume}{331}}, \bibinfo{pages}{439} (\bibinfo{year}{2011}).

\bibitem[{ \citenamefont{Koulakov et~al.}(1996) \citenamefont{Koulakov, Fogler,
 and Shklovskii}}]{koulakov:1996}
\bibinfo{author}{\bibfnamefont{A.~A.} \bibnamefont{Koulakov}},
 \bibinfo{author}{\bibfnamefont{M.~M.} \bibnamefont{Fogler}},
 \bibnamefont{and} \bibinfo{author}{\bibfnamefont{B.~I.}
 \bibnamefont{Shklovskii}}, \emph{\bibinfo{title}{Charge density wave in
 two-dimensional electron liquid in weak magnetic field}},
 \bibinfo{journal}{Phys. Rev. Lett.} \textbf{\bibinfo{volume}{76}},
 \bibinfo{pages}{499} (\bibinfo{year}{1996}).

\bibitem[{ \citenamefont{Fogler et~al.}(1996) \citenamefont{Fogler, Koulakov, and
 Shklovskii}}]{fogler:1996}
\bibinfo{author}{\bibfnamefont{M.~M.} \bibnamefont{Fogler}},
 \bibinfo{author}{\bibfnamefont{A.~A.} \bibnamefont{Koulakov}},
 \bibnamefont{and} \bibinfo{author}{\bibfnamefont{B.~I.}
 \bibnamefont{Shklovskii}}, \emph{\bibinfo{title}{Ground state of a
 two-dimensional electron liquid in a weak magnetic field}},
 \bibinfo{journal}{Phys. Rev. B} \textbf{\bibinfo{volume}{54}},
 \bibinfo{pages}{1853} (\bibinfo{year}{1996}).

\bibitem[{ \citenamefont{Lilly et~al.}(1999{\natexlab{a}}) \citenamefont{Lilly,
 Cooper, Eisenstein, Pfeiffer, and West}}]{lilly:1999a}
\bibinfo{author}{\bibfnamefont{M.~P.} \bibnamefont{Lilly}},
 \bibinfo{author}{\bibfnamefont{K.~B.} \bibnamefont{Cooper}},
 \bibinfo{author}{\bibfnamefont{J.~P.} \bibnamefont{Eisenstein}},
 \bibinfo{author}{\bibfnamefont{L.~N.} \bibnamefont{Pfeiffer}},
 \bibnamefont{and} \bibinfo{author}{\bibfnamefont{K.~W.} \bibnamefont{West}},
 \emph{\bibinfo{title}{Evidence for an anisotropic state of two-dimensional
 electrons in high Landau levels}}, \bibinfo{journal}{Phys. Rev. Lett.}
 \textbf{\bibinfo{volume}{82}}, \bibinfo{pages}{394}
 (\bibinfo{year}{1999}{\natexlab{a}}).

\bibitem[{ \citenamefont{Du et~al.}(1999) \citenamefont{Du, Tsui, Stormer,
 Pfeiffer, Baldwin, and West}}]{du:1999}
\bibinfo{author}{\bibfnamefont{R.~R.} \bibnamefont{Du}},
 \bibinfo{author}{\bibfnamefont{D.~C.} \bibnamefont{Tsui}},
 \bibinfo{author}{\bibfnamefont{H.~L.} \bibnamefont{Stormer}},
 \bibinfo{author}{\bibfnamefont{L.~N.} \bibnamefont{Pfeiffer}},
 \bibinfo{author}{\bibfnamefont{K.~W.} \bibnamefont{Baldwin}},
 \bibnamefont{and} \bibinfo{author}{\bibfnamefont{K.~W.} \bibnamefont{West}},
 \emph{\bibinfo{title}{Strongly anisotropic transport in higher
 two-dimensional Landau levels}}, \bibinfo{journal}{Solid State Commun.}
 \textbf{\bibinfo{volume}{109}}, \bibinfo{pages}{389} (\bibinfo{year}{1999}).

\bibitem[{ \citenamefont{Fradkin and Kivelson}(1999)}]{fradkin:1999}
\bibinfo{author}{\bibfnamefont{E.}~\bibnamefont{Fradkin}} \bibnamefont{and}
 \bibinfo{author}{\bibfnamefont{S.~A.} \bibnamefont{Kivelson}},
 \emph{\bibinfo{title}{Liquid-crystal phases of quantum Hall systems}},
 \bibinfo{journal}{Phys. Rev. B} \textbf{\bibinfo{volume}{59}},
 \bibinfo{pages}{8065} (\bibinfo{year}{1999}).

\bibitem[{ \citenamefont{Fradkin et~al.}(2000) \citenamefont{Fradkin, Kivelson,
 Manousakis, and Nho}}]{fradkin:2000}
\bibinfo{author}{\bibfnamefont{E.}~\bibnamefont{Fradkin}},
 \bibinfo{author}{\bibfnamefont{S.~A.} \bibnamefont{Kivelson}},
 \bibinfo{author}{\bibfnamefont{E.}~\bibnamefont{Manousakis}},
 \bibnamefont{and} \bibinfo{author}{\bibfnamefont{K.}~\bibnamefont{Nho}},
 \emph{\bibinfo{title}{Nematic phase of the two-dimensional electron gas in a
 magnetic field}}, \bibinfo{journal}{Phys. Rev. Lett.}
 \textbf{\bibinfo{volume}{84}}, \bibinfo{pages}{1982} (\bibinfo{year}{2000}).

\bibitem[{ \citenamefont{Zhu et~al.}(2002) \citenamefont{Zhu, Pan, Stormer,
 Pfeiffer, and West}}]{zhu:2002}
\bibinfo{author}{\bibfnamefont{J.}~\bibnamefont{Zhu}},
 \bibinfo{author}{\bibfnamefont{W.}~\bibnamefont{Pan}},
 \bibinfo{author}{\bibfnamefont{H.~L.} \bibnamefont{Stormer}},
 \bibinfo{author}{\bibfnamefont{L.~N.} \bibnamefont{Pfeiffer}},
 \bibnamefont{and} \bibinfo{author}{\bibfnamefont{K.~W.} \bibnamefont{West}},
 \emph{\bibinfo{title}{Density-induced interchange of anisotropy axes at
 half-filled high Landau levels}}, \bibinfo{journal}{Phys. Rev. Lett.}
 \textbf{\bibinfo{volume}{88}}, \bibinfo{pages}{116803}
 (\bibinfo{year}{2002}).

\bibitem[{ \citenamefont{Pollanen et~al.}(2015) \citenamefont{Pollanen, Cooper,
 Brandsen, Eisenstein, Pfeiffer, and West}}]{pollanen:2015}
\bibinfo{author}{\bibfnamefont{J.}~\bibnamefont{Pollanen}},
 \bibinfo{author}{\bibfnamefont{K.~B.} \bibnamefont{Cooper}},
 \bibinfo{author}{\bibfnamefont{S.}~\bibnamefont{Brandsen}},
 \bibinfo{author}{\bibfnamefont{J.~P.} \bibnamefont{Eisenstein}},
 \bibinfo{author}{\bibfnamefont{L.~N.} \bibnamefont{Pfeiffer}},
 \bibnamefont{and} \bibinfo{author}{\bibfnamefont{K.~W.} \bibnamefont{West}},
 \emph{\bibinfo{title}{Heterostructure symmetry and the orientation of the
 quantum Hall nematic phases}}, \bibinfo{journal}{Phys. Rev. B}
 \textbf{\bibinfo{volume}{92}}, \bibinfo{pages}{115410}
 (\bibinfo{year}{2015}).

\bibitem[{ \citenamefont{Liu et~al.}(2013) \citenamefont{Liu, Kamburov, Shayegan,
 Pfeiffer, West, and Baldwin}}]{liu:2013}
\bibinfo{author}{\bibfnamefont{Y.}~\bibnamefont{Liu}},
 \bibinfo{author}{\bibfnamefont{D.}~\bibnamefont{Kamburov}},
 \bibinfo{author}{\bibfnamefont{M.}~\bibnamefont{Shayegan}},
 \bibinfo{author}{\bibfnamefont{L.~N.} \bibnamefont{Pfeiffer}},
 \bibinfo{author}{\bibfnamefont{K.~W.} \bibnamefont{West}}, \bibnamefont{and}
 \bibinfo{author}{\bibfnamefont{K.~W.} \bibnamefont{Baldwin}},
 \emph{\bibinfo{title}{Spin and charge distribution symmetry dependence of
 stripe phases in two-dimensional electron systems confined to wide quantum
 wells}}, \bibinfo{journal}{Phys. Rev. B} \textbf{\bibinfo{volume}{87}},
 \bibinfo{pages}{075314} (\bibinfo{year}{2013}).

\bibitem[{ \citenamefont{{Sodemann} and {MacDonald}}(2013)}]{sodemann:2013}
\bibinfo{author}{\bibfnamefont{I.}~\bibnamefont{{Sodemann}}} \bibnamefont{and}
 \bibinfo{author}{\bibfnamefont{A.~H.} \bibnamefont{{MacDonald}}},
 \emph{\bibinfo{title}{Theory of native orientational pinning in quantum Hall
 nematics}}, \bibinfo{journal}{arXiv:1307.5489} (\bibinfo{year}{2013}).

\bibitem[{ \citenamefont{Koduvayur et~al.}(2011) \citenamefont{Koduvayur,
 Lyanda-Geller, Khlebnikov, Csathy, Manfra, Pfeiffer, West, and
 Rokhinson}}]{kovudayur:2011}
\bibinfo{author}{\bibfnamefont{S.~P.} \bibnamefont{Koduvayur}},
 \bibinfo{author}{\bibfnamefont{Y.}~\bibnamefont{Lyanda-Geller}},
 \bibinfo{author}{\bibfnamefont{S.}~\bibnamefont{Khlebnikov}},
 \bibinfo{author}{\bibfnamefont{G.}~\bibnamefont{Csathy}},
 \bibinfo{author}{\bibfnamefont{M.~J.} \bibnamefont{Manfra}},
 \bibinfo{author}{\bibfnamefont{L.~N.} \bibnamefont{Pfeiffer}},
 \bibinfo{author}{\bibfnamefont{K.~W.} \bibnamefont{West}}, \bibnamefont{and}
 \bibinfo{author}{\bibfnamefont{L.~P.} \bibnamefont{Rokhinson}},
 \emph{\bibinfo{title}{Effect of strain on stripe phases in the quantum Hall
 regime}}, \bibinfo{journal}{Phys. Rev. Lett.} \textbf{\bibinfo{volume}{106}},
 \bibinfo{pages}{016804} (\bibinfo{year}{2011}).

\bibitem[{ \citenamefont{Jungwirth et~al.}(1999) \citenamefont{Jungwirth,
 MacDonald, Smr\ifmmode~\check{c}\else \v{c}\fi{}ka, and
 Girvin}}]{jungwirth:1999}
\bibinfo{author}{\bibfnamefont{T.}~\bibnamefont{Jungwirth}},
 \bibinfo{author}{\bibfnamefont{A.~H.} \bibnamefont{MacDonald}},
 \bibinfo{author}{\bibfnamefont{L.}~\bibnamefont{Smr\ifmmode~\check{c}\else
 \v{c}\fi{}ka}}, \bibnamefont{and} \bibinfo{author}{\bibfnamefont{S.~M.}
 \bibnamefont{Girvin}}, \emph{\bibinfo{title}{Field-tilt anisotropy energy in
 quantum Hall stripe states}}, \bibinfo{journal}{Phys. Rev. B}
 \textbf{\bibinfo{volume}{60}}, \bibinfo{pages}{15574} (\bibinfo{year}{1999}).

\bibitem[{ \citenamefont{Stanescu et~al.}(2000) \citenamefont{Stanescu, Martin,
 and Phillips}}]{stanescu:2000}
\bibinfo{author}{\bibfnamefont{T.~D.} \bibnamefont{Stanescu}},
 \bibinfo{author}{\bibfnamefont{I.}~\bibnamefont{Martin}}, \bibnamefont{and}
 \bibinfo{author}{\bibfnamefont{P.}~\bibnamefont{Phillips}},
 \emph{\bibinfo{title}{Finite-temperature density instability at high Landau
 level occupancy}}, \bibinfo{journal}{Phys. Rev. Lett.}
 \textbf{\bibinfo{volume}{84}}, \bibinfo{pages}{1288} (\bibinfo{year}{2000}).

\bibitem[{ \citenamefont{Lilly et~al.}(1999{\natexlab{b}}) \citenamefont{Lilly,
 Cooper, Eisenstein, Pfeiffer, and West}}]{lilly:1999b}
\bibinfo{author}{\bibfnamefont{M.~P.} \bibnamefont{Lilly}},
 \bibinfo{author}{\bibfnamefont{K.~B.} \bibnamefont{Cooper}},
 \bibinfo{author}{\bibfnamefont{J.~P.} \bibnamefont{Eisenstein}},
 \bibinfo{author}{\bibfnamefont{L.~N.} \bibnamefont{Pfeiffer}},
 \bibnamefont{and} \bibinfo{author}{\bibfnamefont{K.~W.} \bibnamefont{West}},
 \emph{\bibinfo{title}{Anisotropic states of two-dimensional electron systems
 in high Landau levels: Effect of an in-plane magnetic field}},
 \bibinfo{journal}{Phys. Rev. Lett.} \textbf{\bibinfo{volume}{83}},
 \bibinfo{pages}{824} (\bibinfo{year}{1999}{\natexlab{b}}).

\bibitem[{ \citenamefont{Pan et~al.}(1999) \citenamefont{Pan, Du, Stormer, Tsui,
 Pfeiffer, Baldwin, and West}}]{pan:1999}
\bibinfo{author}{\bibfnamefont{W.}~\bibnamefont{Pan}},
 \bibinfo{author}{\bibfnamefont{R.~R.} \bibnamefont{Du}},
 \bibinfo{author}{\bibfnamefont{H.~L.} \bibnamefont{Stormer}},
 \bibinfo{author}{\bibfnamefont{D.~C.} \bibnamefont{Tsui}},
 \bibinfo{author}{\bibfnamefont{L.~N.} \bibnamefont{Pfeiffer}},
 \bibinfo{author}{\bibfnamefont{K.~W.} \bibnamefont{Baldwin}},
 \bibnamefont{and} \bibinfo{author}{\bibfnamefont{K.~W.} \bibnamefont{West}},
 \emph{\bibinfo{title}{Strongly anisotropic electronic transport at Landau
 level filling factor $\nu = 9/2$ and $\nu = 5/2$ under a tilted magnetic field}}, \bibinfo{journal}{Phys.
 Rev. Lett.} \textbf{\bibinfo{volume}{83}}, \bibinfo{pages}{820}
 (\bibinfo{year}{1999}).

\bibitem[{ \citenamefont{Cooper et~al.}(2001) \citenamefont{Cooper, Lilly,
 Eisenstein, Jungwirth, Pfeiffer, and West}}]{cooper:2001}
\bibinfo{author}{\bibfnamefont{K.~B.} \bibnamefont{Cooper}},
 \bibinfo{author}{\bibfnamefont{M.~P.} \bibnamefont{Lilly}},
 \bibinfo{author}{\bibfnamefont{J.~P.} \bibnamefont{Eisenstein}},
 \bibinfo{author}{\bibfnamefont{T.}~\bibnamefont{Jungwirth}},
 \bibinfo{author}{\bibfnamefont{L.~N.} \bibnamefont{Pfeiffer}},
 \bibnamefont{and} \bibinfo{author}{\bibfnamefont{K.~W.} \bibnamefont{West}},
 \emph{\bibinfo{title}{An investigation of orientational symmetry-breaking
 mechanisms in high Landau levels}}, \bibinfo{journal}{Solid State Commun.}
 \textbf{\bibinfo{volume}{119}}, \bibinfo{pages}{89} (\bibinfo{year}{2001}).

\bibitem[{ \citenamefont{Zhu et~al.}(2009) \citenamefont{Zhu, Sambandamurthy,
 Engel, Tsui, Pfeiffer, and West}}]{zhu:2009}
\bibinfo{author}{\bibfnamefont{H.}~\bibnamefont{Zhu}},
 \bibinfo{author}{\bibfnamefont{G.}~\bibnamefont{Sambandamurthy}},
 \bibinfo{author}{\bibfnamefont{L.~W.} \bibnamefont{Engel}},
 \bibinfo{author}{\bibfnamefont{D.~C.} \bibnamefont{Tsui}},
 \bibinfo{author}{\bibfnamefont{L.~N.} \bibnamefont{Pfeiffer}},
 \bibnamefont{and} \bibinfo{author}{\bibfnamefont{K.~W.} \bibnamefont{West}},
 \emph{\bibinfo{title}{Pinning mode resonances of 2d electron stripe phases:
 Effect of an in-plane magnetic field}}, \bibinfo{journal}{Phys. Rev. Lett.}
 \textbf{\bibinfo{volume}{102}}, \bibinfo{pages}{136804}
 (\bibinfo{year}{2009}).

\bibitem[{ \citenamefont{Shi et~al.}(2016) \citenamefont{Shi, Zudov, Watson,
 Gardner, and Manfra}}]{shi:2016b}
\bibinfo{author}{\bibfnamefont{Q.}~\bibnamefont{Shi}},
 \bibinfo{author}{\bibfnamefont{M.~A.} \bibnamefont{Zudov}},
 \bibinfo{author}{\bibfnamefont{J.~D.} \bibnamefont{Watson}},
 \bibinfo{author}{\bibfnamefont{G.~C.} \bibnamefont{Gardner}},
 \bibnamefont{and} \bibinfo{author}{\bibfnamefont{M.~J.}
 \bibnamefont{Manfra}}, \emph{\bibinfo{title}{Reorientation of quantum Hall
 stripes within a partially filled Landau level}}, \bibinfo{journal}{Phys.
 Rev. B} \textbf{\bibinfo{volume}{93}}, \bibinfo{pages}{121404(R)}
 (\bibinfo{year}{2016}).

\bibitem[{ \citenamefont{Xia et~al.}(2010) \citenamefont{Xia, Cvicek, Eisenstein,
 Pfeiffer, and West}}]{xia:2010}
\bibinfo{author}{\bibfnamefont{J.}~\bibnamefont{Xia}},
 \bibinfo{author}{\bibfnamefont{V.}~\bibnamefont{Cvicek}},
 \bibinfo{author}{\bibfnamefont{J.~P.} \bibnamefont{Eisenstein}},
 \bibinfo{author}{\bibfnamefont{L.~N.} \bibnamefont{Pfeiffer}},
 \bibnamefont{and} \bibinfo{author}{\bibfnamefont{K.~W.} \bibnamefont{West}},
 \emph{\bibinfo{title}{Tilt-induced anisotropic to isotropic phase transition
 at $\ensuremath{\nu}=5/2$}}, \bibinfo{journal}{Phys. Rev. Lett.}
 \textbf{\bibinfo{volume}{105}}, \bibinfo{pages}{176807}
 (\bibinfo{year}{2010}).

\bibitem[{ \citenamefont{Xia et~al.}(2011) \citenamefont{Xia, Eisenstein,
 Pfeiffer, and West}}]{xia:2011}
\bibinfo{author}{\bibfnamefont{J.}~\bibnamefont{Xia}},
 \bibinfo{author}{\bibfnamefont{J.~P.} \bibnamefont{Eisenstein}},
 \bibinfo{author}{\bibfnamefont{L.~N.} \bibnamefont{Pfeiffer}},
 \bibnamefont{and} \bibinfo{author}{\bibfnamefont{K.~W.} \bibnamefont{West}},
 \emph{\bibinfo{title}{Evidence for a fractionally quantized Hall state with
 anisotropic longitudinal transport}}, \bibinfo{journal}{Nat. Phys.}
 \textbf{\bibinfo{volume}{7}}, \bibinfo{pages}{845} (\bibinfo{year}{2011}).

\bibitem[{ \citenamefont{Friess et~al.}(2014) \citenamefont{Friess, Umansky,
 Tiemann, von Klitzing, and Smet}}]{friess:2014}
\bibinfo{author}{\bibfnamefont{B.}~\bibnamefont{Friess}},
 \bibinfo{author}{\bibfnamefont{V.}~\bibnamefont{Umansky}},
 \bibinfo{author}{\bibfnamefont{L.}~\bibnamefont{Tiemann}},
 \bibinfo{author}{\bibfnamefont{K.}~\bibnamefont{von Klitzing}},
 \bibnamefont{and} \bibinfo{author}{\bibfnamefont{J.~H.} \bibnamefont{Smet}},
 \emph{\bibinfo{title}{Probing the microscopic structure of the stripe phase
 at filling factor $5/2$}}, \bibinfo{journal}{Phys. Rev. Lett.}
 \textbf{\bibinfo{volume}{113}}, \bibinfo{pages}{076803}
 (\bibinfo{year}{2014}).

\bibitem[{ \citenamefont{Pan et~al.}(2000) \citenamefont{Pan, Jungwirth, Stormer,
 Tsui, MacDonald, Girvin, Smrc\ifmmode~\breve{}\else \u{}\fi{}ka, Pfeiffer,
 Baldwin, and West}}]{pan:2000}
\bibinfo{author}{\bibfnamefont{W.}~\bibnamefont{Pan}},
 \bibinfo{author}{\bibfnamefont{T.}~\bibnamefont{Jungwirth}},
 \bibinfo{author}{\bibfnamefont{H.~L.} \bibnamefont{Stormer}},
 \bibinfo{author}{\bibfnamefont{D.~C.} \bibnamefont{Tsui}},
 \bibinfo{author}{\bibfnamefont{A.~H.} \bibnamefont{MacDonald}},
 \bibinfo{author}{\bibfnamefont{S.~M.} \bibnamefont{Girvin}},
 \bibinfo{author}{\bibfnamefont{L.}~\bibnamefont{Smrc\ifmmode~\breve{}\else
 \u{}\fi{}ka}}, \bibinfo{author}{\bibfnamefont{L.~N.} \bibnamefont{Pfeiffer}},
 \bibinfo{author}{\bibfnamefont{K.~W.} \bibnamefont{Baldwin}},
 \bibnamefont{and} \bibinfo{author}{\bibfnamefont{K.~W.} \bibnamefont{West}},
 \emph{\bibinfo{title}{Reorientation of anisotropy in a square well quantum
 Hall sample}}, \bibinfo{journal}{Phys. Rev. Lett.}
 \textbf{\bibinfo{volume}{85}}, \bibinfo{pages}{3257} (\bibinfo{year}{2000}).

\bibitem[{not({\natexlab{a}})}]{note:90}
\bibinfo{note}{Qualitatively consistent with our observations, early
 experiments \cite{lilly:1999b} have found that, when $\bip$ is applied
 perpendicular to stripes, the anisotropy is affected more at $\nu$ = 9/2 and
 13/2 than at $\nu$ = 11/2 and 15/2.}

\bibitem[{not({\natexlab{b}})}]{note:8}
\bibinfo{note}{An alternative senario is that, the effect of $\bip$ is modified
 by disorder which is anisotropic along the two cystal axes \cite{zhu:2009}.
 However, effect of disorder on stripes orientation has not been
 studied theoretically and it seems unlikely that such a mechanism would
 depend on $\sigma$.}

\bibitem[{not({\natexlab{c}})}]{note:9}
\bibinfo{note}{We note that the anisotropy energy due to this mechanism also
 shows oscillating behavior with $\sigma$.}

\bibitem[{not({\natexlab{d}})}]{note:15}
\bibinfo{note}{Recent experiments \cite{pollanen:2015} have found that the
 quantum well symmetry has a minor role in deciding the stripes orientation,
 suggesting that spin-orbital interaction \cite{sodemann:2013} is unlikely to
 be the major mechanism for the native symmetry breaking.}

\end{thebibliography}

\end{document}